\documentclass[a4paper,11pt]{article}
\usepackage{pos}

\def\Z{\mathbb{Z}}
\def\R{\mathbb{R}}

\title{Generalised Symmetries in Particle Physics}

\author*[a]{Joe Davighi}

\affiliation[a]{CERN,\\
   1, Esplanade des Particules, Meyrin, Switzerland}

\emailAdd{joseph.davighi@cern.ch}

\abstract{In this talk I review various notions of generalised global symmetry: higher-form, higher-group, and non-invertible symmetry. All these notions have had profound impact on quantum field theory research in the last decade. I highlight various applications of these new symmetries in particle physics, focussing on theories beyond the Standard Model. Areas touched upon include axions, gauge unification, dark matter, neutrino masses, and flavour hierarchies.
}

\FullConference{9th Symposium on Prospects in the Physics of Discrete Symmetries (DISCRETE2024)\\
 2–6 Dec 2024\\
Ljubljana, Slovenia\\}

\begin{document}
\maketitle

\section{Introduction}

The notion of “global symmetry” has been generalised in various directions in the last decade following the now-seminal work~\cite{Gaiotto:2014kfa}, which has had a significant impact on formal quantum field theory research in that time. 
In this talk, we discuss what can we can learn about particle physics from generalised symmetries. This will not be a comprehensive review by any means. Rather, my limited aims are:
(i) to introduce the various kinds of generalised symmetry studied in formal theory, primarily for an audience of particle physicists, and
(ii) to select a few examples that illustrate things people have tried to do with generalised symmetries in particle physics so far -- needless to say, my selection is far from being impartial.
I refer the interested reader to 
a number of excellent reviews and lecture notes that have been recently written on the topic: including Refs.~\cite{Gomes:2023ahz,Schafer-Nameki:2023jdn,Bhardwaj:2023kri,Shao:2023gho,Iqbal:2024pee} from a more formal perspective, and Refs.~\cite{Brennan:2023mmt,Reece:2023czb}
from a more phenomenological perspective.

The plan of this talk is as follows. To bring readers up to speed on the basic pre-requisites, in \S\ref{sec:SDO} I begin by recasting the familiar notion of global symmetry in quantum field theory in terms of topological defects.
In the remaining Sections I introduce three kinds of generalised symmetry, each accompanied by a select example or two where this symmetry type has been applied in a particle physics context.
In \S\ref{sec:nform} I discuss higher-form symmetry (the main focus of this talk), 
before generalising this to higher-group symmetry in \S\ref{sec:ngroup}. 
In \S\ref{sec:NIS} I briefly discuss non-invertible symmetry.

\section{Symmetries as Topological Defects} \label{sec:SDO}

We are used to deriving symmetries from an action. For example, consider the theory of a complex scalar field $\phi$ with Lagrangian $L=\partial \phi \partial \phi^\dagger - V(\phi \phi^\dagger)$. This Lagrangian is invariant under the transformation $\phi(x) \mapsto e^{i\alpha}\phi(x)$, where the parameter $\alpha \in 2\pi \R/\Z\cong U(1)$. For such a continuous global symmetry, we can derive a conserved current $j^\mu$ from the Lagrangian using Noether's procedure, {\em viz.} $j^\mu = \delta \phi \cdot \frac{\partial L}{\partial (\partial_\mu \phi)}$, which satisfies the conservation law $\partial_\mu j^\mu=0$ on the classical equations of motion (the Euler--Lagrange equations). 
In our example, $j^\mu \sim i[(\partial^\mu \phi^\dagger)\phi-\phi^\dagger (\partial^\mu \phi)]$. 
We proceed to construct a conserved charge by integrating. In a Lorentzian theory this might be defined as $Q(t)\sim \int_{M_{d-1}^t} d^{d-1}x \, j^0$ where $M_{d-1}^t$ is a spatial slice at time $t$.
This charge is conserved in time up to a boundary term, $\dot{Q}(t)=\int_{M_{d-1}^t}d^{d-1}x\, \partial_t j^0=-\int_{\partial M_{d-1}^t} d^{d-2}x\, \Vec{n} \cdot \Vec{j}$, using the divergence theorem. Finally, one can couple this theory to a gauge field $A_\mu$ for the $U(1)$ symmetry via $S=\int_{\Sigma_d}A_\mu j^\mu$, and promote the global symmetry to a local one; in our example, $\phi\mapsto e^{i\alpha(x)}\phi$, $A_\mu \to A_\mu +\partial_\mu \alpha$ where $\alpha(x):\Sigma_d\to U(1)$ is some smooth map.

All this can be rephrased using the language of differential forms, where recall that
a differential $k$-form on manifold $\Sigma_d$ is a totally antisymmetric tensor of type $(0,k)$. Given local coordinates $x^\mu$ on $\Sigma_d$, one can expand the $k$-form in terms of the coordinate basis 1-forms $dx^\mu$, as $A=\frac{1}{k!}A_{[\mu_1\dots \mu_k]}dx^{\mu_1}\wedge \dots \wedge dx^{\mu_k}$, where $\wedge$ denotes the wedge product.
The components of the conserved current $j^\mu$ can be used to define a $(d-1)$-form $\star j=\epsilon_{\mu_1\mu_2\dots\mu_{d}} j^{\mu_1} dx^{\mu_2} \wedge \dots \wedge dx^{\mu_d}$, where $\epsilon$ is the completely antisymmetric Levi-Civita tensor -- we choose to put the `$\star$' in front because it is more typical to define the current ($j$) as a 1-form which can be obtained from the above by dualising. The continuity equation translates to $d\star j=0$, where $d$ is the exterior derivative. 
Charge operators $Q(M_{d-1})=\int_{M_{d-1}}\star j$ can naturally be defined on any codimension-1 submanifold $M_{d-1}$, and here it is convenient to imagine we have Wick-rotated to Euclidean signature so that there is no preferred time-direction.
Closure of $\star j$ then means that this charge is a {\em topological operator} in the sense that it does not depend on the homology class of $M_{d-1}$; being very cavalier with our notation, we have $Q(M_{d-1}+\partial Y_d) = Q(M_{d-1})$. Colloquially, this tells us that the measured charge is not changed under `small wiggles' of the surface $M_{d-1}$. Finally, the gauge field $A_\mu$ naturally defines a 1-form $A=A_\mu dx^\mu$, and the minimal coupling term is $\int_{\Sigma_d}A\wedge \star j$.

The charge operators $Q(M_{d-1})$ that we have defined, obtained from the infinitesimal Noether procedure, live in the Lie algebra $\mathfrak{g}$ of the symmetry group $G$ (where $G=U(1)$, $\mathfrak{g}=\R$ in our example). One can exponentiate these charge operators to get group-valued topological operators, 
\begin{equation} \label{eq:U1SDO}
    U_{g=e^{i\alpha}} (M_{d-1}) := \exp i\alpha Q(M_{d-1})=\exp i\alpha \int_{M_{d-1}} \star j\, .
\end{equation}
In the Lorentzian picture introduced above whereby $M_{d-1}^t$ was taken to be a spatial slice at fixed time $t$, these group valued operators have the interpretation of being unitary operators $U_g(t)$ acting on the Hilbert space $\mathcal{H}_t$. Being symmetries means these unitary operators commute with the Hamiltonian operator that generates time evolution -- the familiar picture from elementary quantum mechanics.

In our more general picture where we evaluate the charge on any codimension-1 manifold,
the topological operator \eqref{eq:U1SDO} acts on local operators $\mathcal{O}(x)$ as $U_g(M_{d-1})O(x)=e^{i\alpha q_{\mathcal{O}}} \mathcal{O}(x)$ if the point $x$ is inside $M_{d-1}$, by which we mean $x$ is {\em linked} by the codimension-1 manifold $M_{d-1}$, and $U_g$ acts trivially on $\mathcal{O}(x)$ if $x$ does not link $M_{d-1}$.
These operators possess a number of key properties:
\begin{enumerate}
    \item The $U_g(M_{d-1})$ are all topological, in this case because $d\star j=0$.
    \item The algebra formed by the set of these operators has a group structure, $g\in G$. This can be inferred by sequentially applying the definition of the action of $U_g$ on local operators.
    \item The $U_g(M_{d-1})$ act on local operators; codimension-1 surfaces link with points.
\end{enumerate}
This suggests an abstract, action-free definition of symmetry directly in terms of the set of topological operators satisfying these properties.
We emphasize that for continuous symmetries, the key property of being {\em topological} was guaranteed by the existence of a current $\star j$, which is a closed, differential-form valued operator of degree $d-1$ (or, if dualised, degree-1). 
But this definition of symmetry naturally works for discrete symmetries too, for which there is no continuous current; rather, one can directly define the topological operators $U_g(M_{d-1})$ for {\em e.g.} $G=\Z_n$ by their action on charged operators, $U_{p\, \text{mod}\, n}(M_{d-1}):\mathcal{O}(x)\mapsto e^{2\pi i\frac{p}{n} q_{\mathcal{O}}}\mathcal{O}(x)$ if $M_{d-1}$ links $x$.

This definition of symmetry in terms of topological operators suggests natural generalisations. The idea is to retain the topological property (1) as the defining property of a symmetry, but to relax properties (2) and (3) above. Relaxing (2), we allow symmetry operators to furnish a more general structure than a group -- giving rise to so-called {\em non-invertible symmetry}. Relaxing (3), we consider topological operators that link not only points, but link higher dimension submanifolds and extended operators defined thereon. Linking symmetry generators with extended objects gives the notion of {\em higher-form symmetry}, which we discuss first.

\section{Higher-form symmetries} \label{sec:nform}

\begin{figure}
    \centering
    \includegraphics[width=0.7\linewidth]{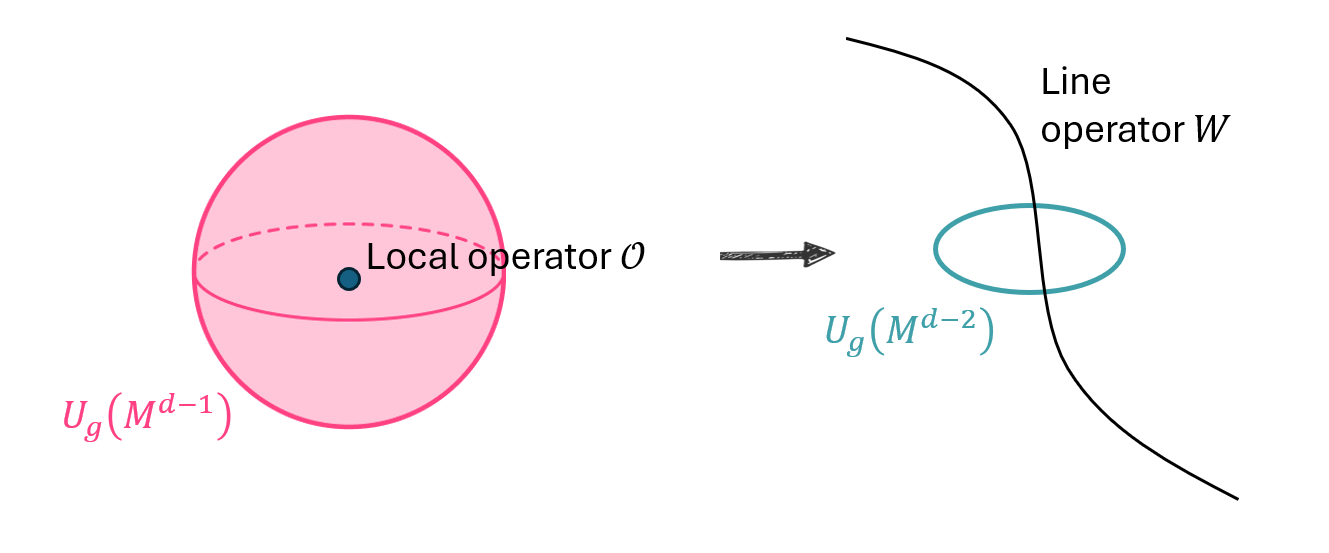}
    \caption{Ordinary symmetries correspond to topological operators that link with points, which can be used to define their action on local operators (left). 1-form symmetries correspond to topological operators in one dimension fewer that link with lines, and so act on extended line operators such as Wilson lines (right). }
    \label{fig:1form}
\end{figure}

We have seen that ordinary (henceforth “0-form”) symmetries act on local operators; the objects charged under an ordinary symmetry are 0-dimensional.
Generalising this, we postulate that a theory might also exhibit 1-form symmetries, defined by topological operators that link with lines, and so act on charged objects that are 1-dimensional line operators  -- the basic idea is illustrated in Fig.~\ref{fig:1form}.
Ringing the changes, the conserved current $J=\star j^{(2)}$ that we integrate (then exponentiate) to obtain the topological operator is now a closed $(d-2)$-form, assuming we have a {\em continuous} 1-form symmetry; the background gauge field is now a 2-form field $B$ whose curvature is a closed 3-form, with background gauge transformation $B^{(2)} \mapsto B^{(2)}+d\Lambda^{(1)}$; and the minimal coupling term added to gauge the 1-form symmetry is $S=\int_{M_d} B^{(2)} \wedge \star j^{(2)}$. 

For example, in the case of a $U(1)$-valued 1-form symmetry, there will be a current with antisymmetric components $j^{\mu\nu}$ satisfying $\partial_\mu j^{\mu\nu}=0$, equivalently $d\star j^{(2)}=0$ with $\star j^{(2)}$ defined analogously to before. This closure property (the continuity equation, when expressed in components) means that the following group-valued operators $U_{e^{i\alpha}}(M_{d-2})=\exp\left(i\alpha \int_{M_{d-2}}\star j^{(2)}\right)$ are topological, and so define symmetries.

We can continue to define a general higher $p$-form symmetry, for which the charged objects are extended $p$-dimensional operators; the current that we integrate is now $J=\star j^{(p+1)}$, which is a closed $(d-(p+1))$-form. 
Note that being a {\em form} means that, if we expand either $J$ or the dualised $j$ using a coordinate basis, then the components are always totally antisymmetric upon exchanging indices.

Higher-form symmetries are always abelian~\cite{Gaiotto:2014kfa}. This is because any two surfaces $M_{k\leq d-2}$ and $M^\prime_{k\leq d-2}$ can be freely moved around eachother in the ambient space $\Sigma_d$ and so there is no well-defined notion of operator ordering except for 0-form symmetries. This means that any pair of higher-form symmetries $U_g(M_{k\leq d-2})$ and $U_{g^\prime}(M^\prime_{k\leq d-2})$ must commute.

\subsection{How to find them?}

It is clear that higher-form symmetries cannot be deduced from variation of the action: the Lagrangian density is a local operator, and so its variation naturally yields a 1-form (equivalently a $(d-1)$-form). So, given a particular quantum field theory, how do we find these higher-form symmetries? The idea is to bypass the action and jump straight to identifying the Noether current operator -- at least in the case of a continuous symmetry.

\subsubsection{Continuous higher-form symmetries}

To identify that a theory possesses a continuous $p$-form symmetry, we need to identify a $p$-form-valued operator $j^{(p)}$ in the theory that is `co-closed', {\em i.e.} satisfies $d\star j^{(p)}=0$, for $p>1$. Equivalently we can find the closed $(d-p)$-form $J=\star j^{(p)}$ directly.

\paragraph{Example: 4d Maxwell.}
This is best illustrated by the simple example of Maxwell theory, that describes free photons in 4d. The action is $S=\int_{\Sigma_4} da\wedge \star da$, where $a$ is the abelian gauge field.
This theory has two closed 2-form-valued operators, 
\begin{equation}
    j^{e}:=f=da \qquad \text{and} \qquad j^m:=\star f\, .
\end{equation}
Conservation is the closure condition: for the `electric 1-form symmetry' this is the Gauss law equation $d\star j^e=d\star f=0$ which holds because there is no electrically charged matter; while for the `magnetic 1-form symmetry' conservation holds by the Bianchi identity $df=0$ (equivalently, the condition that there are no magnetic monopoles). We can define two sets of topological operators. For the electric 1-form symmetry these are 
\begin{equation} \label{eq:electric}
    U_g^e(M_2)=\exp\left(i\alpha \int_{M_2}\star f\right)= \exp\left(i\alpha \int_{M_2}\Vec{E}\cdot \Vec{dS}\right)\, , 
\end{equation}
while for the magnetic case simply replace $\star f$ by $f$ in the second expression, and $\Vec{E}$ by $\Vec{B}$ in the third.
These act on {\em i.e.} link with Wilson line operators 
\begin{equation}
    L_q(\gamma)=\exp i q\oint_\gamma A\, ,
\end{equation}
which are worldlines of non-dynamical heavy charges (or on 't Hooft lines, the dualised version, in the magnetic case), by
\begin{equation} \label{eq:Wilson}
    U_{e^{i\alpha}}^e(M_2) \cdot L_q(\gamma) = e^{iq\alpha \text{Link}(M_2,\gamma)}L_q(\gamma)
\end{equation}
The action of the magnetic 1-form symmetry operators on 't Hooft lines is analogous.

\subsubsection{Discrete higher-form symmetries}

A {\em non}-example of a theory with continuous 1-form symmetry is Yang--Mills with gauge group $SU(N)$. In going from $U(1)$ to $SU(N)$, the Maxwell equation $d\star f=0$ is replaced by the non-abelian version, $D\star f=(d+a\wedge)f=0$, and so we lose the gauge-invariant closed 2-form and so the continuous electric 1-form symmetry.

But that is not the end of the story -- Yang--Mills instead has {\em discrete} 1-form symmetry. How does one find discrete higher-form symmetries? Here we again cannot start from the action, but nor can we even start from a current! Rather, we should look directly for the charged operators which the higher-form symmetry acts on. One finds that $SU(N)$ Yang--Mills, with action $S\sim \int_{\Sigma_d}\text{Tr} f\wedge \star f$, has a $\Z_N$-valued 1-form symmetry that acts on Wilson lines (say, in the fundamental representation of $SU(N)$) which cannot be screened by local operators.

\paragraph{Example: Standard Model gauge group.}
The discrete 1-form symmetry distinguishes the global form of otherwise identical gauge groups~\cite{Aharony:2013hda}. For instance, Yang--Mills with $G=SU(N)$ has an electric $\Z_N$ 1-form symmetry (acting on Wilson lines), whereas Yang--Mills with gauge group $G=SU(N)/\Z_N$, where the quotient by the centre $\Z_N$ subgroup precludes electrically-charged matter in the fundamental representation, has magnetic $\Z_N$ 1-form symmetry (acting on 't Hooft lines). Of closer relevance to particle physics, there are four different versions of the Standard Model gauge group~\cite{Tong:2017oea}, assuming it is connected, that are all compatible with the SM matter content, but which have different 1-form symmetries and corresponding spectra of line operators. 

We note in passing that this 1-form `centre symmetry' in 4d Yang--Mills theory has had important applications in formal theory, especially when combined with the idea of anomalies. In~\cite{Gaiotto:2017yup} it was shown that 4d Yang--Mills with gauge group $G=SU(2N)$ defined at $\theta=\pi$, where the action includes the theta term $S_\theta = \frac{\theta}{8\pi^2}\int \text{Tr~} f\wedge f$, which classically preserves parity symmetry, has a mixed anomaly involving the $\Z_{2N}$ electric 1-form symmetry and parity. The need to match this anomaly was used to prove the vacuum of this theory at $\theta=\pi$ must be non-trivial.

\subsection{How to break them?}

It is `harder' to break higher-form symmetries than it is to break ordinary symmetries.
This is because they do not `see' local operators in the Lagrangian, meaning they cannot be broken by perturbing the theory through the inclusion of small irrelevant operators, as we are used to for 0-form accidental symmetries in the context of an effective field theory.

Instead, higher-form symmetries are broken by introducing new degrees of freedom.

\paragraph{Example: Maxwell $\to$ QED.}
Inclusion of electrically-charged matter modifies the Maxwell equation to $d\star j^e=d\star f=\rho_e(x)\neq 0$, so we lose closure of the current and thus lose the 1-form symmetry: as if familiar from electromagnetism, the Gaussian surface operators defined in \eqref{eq:electric} will no longer be topological in the presence of dynamical charged matter.

\paragraph{Example: Axion quality from 1-form symmetry.}

We now turn to an example of relevance to BSM physics~\cite{Craig:2024dnl}.
Consider Minkowski spacetime extended by one extra compact dimension, $M_5=\R^{1,3}\times S^1_R$, coupled to a 5d $U(1)$ gauge field $c$. The 5d action is
\begin{equation}
    S=\int_{M_5} -\frac{1}{2g^2} dc\wedge \star (dc) + \frac{N}{8\pi^2}c\wedge \mathrm{Tr~}(G\wedge G)
\end{equation}
where $G$ is the gluon field extended to $M_5$. In this extra-dimensional setup, the {\em axion} is the zero-mode of the 5d gauge field along the circle. It is simply a Wilson line of the extra-dimensional theory:
\begin{equation}
    a=\int_{S^1} c\, ,
\end{equation}
which is naturally periodic, with $a\sim a+2\pi f$ where $f=(g\sqrt{2\pi R})^{-1}$.
Recall that 1-form symmetries act on line operators in the theory; in particular, given the abelian gauge field $c$ there is a $U(1)$-valued electric 1-form symmetry acting on the Wilson lines $L_q(\gamma)=\exp iq\oint_\gamma c$, as
\begin{equation}
    U_{e^{i\alpha}}^e(M_{3}) \cdot L_q(\gamma) = e^{iq\alpha \text{Link}(M_3,\gamma)}L_q(\gamma)\, ,
\end{equation}
by \eqref{eq:Wilson}. Note that because we are in 5 dimensions, the 1-form symmetry operator is obtained by integration over a 3-manifold $M_3$. 
Wrapping $\gamma$ around the $S^1_R$ and then dimensionally reducing the theory on this circle, this becomes simply a shift of the axion field:
\begin{equation}
    a \mapsto a + \alpha f\, .
\end{equation}
So, we learn that the axion shift symmetry is equivalent, in the 5d picture, to an electric 1-form symmetry.

This in turn gives insight into the generation of a {\em potential} for the axion: since any non-flat potential explicitly breaks the axion shift symmetry, we should also be able to understand any non-flat axion potential as being generated by explicit breaking of the 5d 1-form symmetry.
But we just saw that higher-form symmetries are in a sense harder to break. 
By considering all possible ways to break the 1-form symmetry, which is a shorter list than classifying all irrelevant local operators that break a shift symmetry, one can classify all the inequivalent ways to generate an axion potential. They are~\cite{Craig:2024dnl}: (i) via electrically-charged matter (in 5d); (ii) via gauging of a magnetic 2-form symmetry; (iii) via gauging of the electric 1-form symmetric; and (iv) via an ABJ anomaly.

While this 5d axion story is in a sense already known, it is an example of a general statement: a $p$-form symmetry in $(4+p)$-dimensions gives rise to a $0$-form symmetry in 4d upon dimensional reduction.
This is well known also in string theory. But it might be interesting to explore if there are other particle physics applications of this simple statement.

\paragraph{Example: higher-form global symmetries from discrete gauge symmetries.}

Discrete symmetries in 4d are often invoked in model building, for example to explain the structure of neutrino masses and mixing angles~\cite{King:2013eh}.
If such discrete symmetries are gauged, as would occur if the remnant of a spontaneously broken continuous gauge symmetry (or
which we might enforce if we wish them to give exact selection rules~\cite{Davighi:2022qgb,Koren:2022bam,Koren:2022axd}), this generically leads to higher-form discrete {\em global} symmetries.
This is because discrete gauge fields typically allow for topologically non-trivial extended operators, such as strings, which cannot be shrunk away because they carry a quantized topological charge.\footnote{The classification of such topologically non-trivial extended operators can be formalised using (reduced) {\em bordism} groups; for a discrete gauge group $N$, the possible such `unshrinkable' objects, of dimension $q$, are classified by the (reduced) bordism groups $\widetilde{\Omega}_q^{\text{Spin}}(BN)$, where $BN$ is the classifying space of $N$ and where we assume the theory (which features fermions) is defined using a spin structure. For example, take $N=A_4$, the group of even permutations on four elements -- a popular choice in neutrino mass model building. The non-trivial bordism group $\widetilde\Omega_1(BA_4) = \Z_3$~\cite{Davighi:2022icj} associated with the abelianization of $A_4$ implies the existence of topological defect operators that are circles with $\Z_3$ holonomy; these link with surfaces in 4d, and so generate a 2-form global symmetry. These discussions are based on work in progress with Markus Dierigl.
\label{foot:discrete}
} 
A higher-form symmetry measures this charge.

One reason this could be important in particle physics is that quantum gravity tells us there ought to be no global symmetries in a fundamental theory (see~\cite{Reece:2023czb} for a phenomenologist-oriented review, and references therein). 
The colloquial argument is that, if one were to throw one of these topologically stable defects into a black hole, there is no way to radiate the topological charge~\cite{Zeldovich:1976vq,Banks:2010zn}.
This means the fundamental theory must break (or gauge) the higher-form symmetry associated with the discrete gauge symmetry. 
We saw this was hard for a higher-form symmetry, but that it can be done via dynamical extended objects; for instance,  the 2-form symmetry discussed in footnote~\ref{foot:discrete}, can be killed if there are dynamical {\em strings} that screen the charged surface operators.
This argument applies the essence of the `swampland cobordism conjecture'~\cite{McNamara:2019rup} to particle physics theories with discrete gauge symmetries.

\section{Higher-group symmetries} \label{sec:ngroup}

We have seen how  higher-form symmetries are generated by topological operators of codimension greater than one that act on extended operators. A further generalisation follows from the realisation that higher form symmetries of different degrees can mix, to form what is known as a `higher-group' structure in mathematics (described by higher-bundles with connection). 
The simplest case is 2-group symmetry~\cite{Kapustin:2013uxa,Sharpe:2015mja,Cordova:2018cvg}, whereby a 0-form symmetry and a 1-form symmetry are intertwined.
One way to understand this intertwining is via the transformation laws for the background gauge fields, in this case a 1-form gauge field $A^{(1)}$ for the 0-form symmetry, and a 2-form gauge field $B^{(2)}$ for the 1-form symmetry. 
In the case that both are $U(1)$-valued symmetries, the 2-group gauge transformation law is
\begin{align} \label{eq:2group}
    A^{(1)} &\mapsto A^{(1)} + d\alpha \, , \\    
    B^{(2)} &\mapsto B^{(2)} + d\Lambda^{(1)} + \frac{n}{2\pi} \alpha dA^{(1)}\, .
\end{align}
Here $n\in \Z$ is called the {\em Postnikov class}, and classifies the particular 2-group symmetry we have (having already fixed the 0-form and 1-form symmetry groups to each be $U(1)$). We will see field theory interpretations of this integer-valued class shortly. 
A generalisation involving higher-form symmetries up to $n$-form symmetry is described by an $n$-group symmetry structure, the mathematical description of which is rather involved. We limit ourselves to considering 2-groups here.

Physically, the inference of 2-group structure is not just an artefact of turning on background fields; 
especially important,
the 2-group results in a modification of the current algebra. Suppose a theory has 0-form symmetry group $G^{(0)}$, whose Lie algebra has structure constants $f_{abc}$, and (abelian) 1-form symmetry group $H^{(1)}=U(1)$. 
The familiar Ward identities for the 0-form currents $j_a^{(1)}$ become `twisted' by the 2-form current $j^{(2)}$~\cite{Cordova:2018cvg,Cordova:2020tij}, for example:
\begin{equation}
    i\partial_\mu j_a^{(1)\,\mu}(x)j_b^{(1)\,\nu}(y) + f_{abc}\delta(x-y) j_c^{(1)\,\nu}(y)=n\frac{1}{8\pi^2} \delta_{ab}\partial_\rho \delta(x-y) j^{(2)\, \rho\nu}(y)\, .
\end{equation}
Because the Postnikov class $n$ appearing on the right-hand side is an integer, it cannot change continuously under any deformation, including the renormalisation group (RG) flow of the theory with energy. Hence, just like an anomaly, the 2-group structure must match from the ultraviolet to the infrared, or else be completely broken. 

This rigidity bestows the 2-group structure with more power than 1-form and 0-form symmetries in isolation: an RG flow cannot break only the 1-form symmetry without explicitly breaking the 0-form “flavour” symmetry at the same scale. This was named the `2-group emergence theorem' in~\cite{Cordova:2018cvg,Cordova:2020tij}.

\paragraph{Example: Unification.}
This 2-group emergence theorem was used in~\cite{Cordova:2022qtz}, together with the observation of 2-group structure in the SM (in the limit of zero Yukawa couplings) that mix $U(3)^5$ flavour symmetries with the 1-form symmetry associated to weak hypercharge, to study the possible embeddings of the SM gauge group(s) inside semi-simple groups.

\paragraph{Example: Topological portal to the dark sector.}
In~\cite{Davighi:2024zip} a new portal to dark matter was proposed, that is a topological effective interaction. In~\cite{Davighi:2024zjp} it was then shown that this portal encodes a 2-group structure, that mixes the non-abelian 0-form chiral flavour symmetries of QCD with a 1-form symmetry acting on the dark sector degrees of freedom. We review how this works in a little detail here.

The topological portal is an operator appearing in the effective theory of QCD pions coupled to a sector of dark pNGBs living on a 2-sphere $S^2\cong SU(2)_D/U(1)_D$ that we suppose arises from some spontaneously broken global symmetry, which we suppose is valid at energies $\sim\mathcal{O}(0.1\div 1)$ GeV. The action for the topological portal is defined, like the Wess--Zumino--Witten term in QCD~\cite{Wess:1971yu,Witten:1983tw}, by extending spacetime to a bulk 5-manifold and integrating a closed, invariant 5-form thereon:
\begin{equation} \label{eq:topDM}
	S[\Sigma_d] = n\int_{X_5} \frac{1}{24\pi^2}\mathrm{Tr}(g^{-1}dg)^3 \wedge \mathrm{Vol}_{S^2}\, , \qquad \partial X_5=\Sigma_4\, ,
\end{equation}
where $g(x)=\exp(i\pi_a(x)t_a/f_\pi)\,:\,X_5\to SU(3)$ is the matrix-valued pion field of QCD, and $\mathrm{Vol}_{S^2}$ is the volume form on the dark $S^2$ factor. In analogy with the WZW term, the coefficient $n\in \Z$ must be quantised for this to define a local ({\em i.e.} bulk-independent) 4d quantum field theory~\cite{Witten:1983tw,Alvarez:1984es,Davighi:2018inx}.

This EFT can deliver an elegant realisation of light thermal inelastic dark matter~\cite{Tucker-Smith:2001myb,Izaguirre:2015zva}. Upon gauging QED (regarded as a subgroup of the QCD flavour symmetry) one obtains from \eqref{eq:topDM} an interaction involving a photon, 
\begin{equation} \label{eq:topGauge}
    S\sim n\int_{\Sigma_4} \frac{e}{8\pi^2 f_\pi f_D^2} \pi_0 F\wedge d\chi_1\wedge d\chi_2\, ,
\end{equation}
where $\chi_{1,2}$ are local coordinates on the dark $S^2$ {\em i.e.} the pair of dark matter fields, and $f_D$ is the scale of dark sector symmetry breaking. This $2 \leftrightarrow 2$ interaction can explain the observed dark matter relic abundance via freeze-out for dark matter masses up to a few GeV. At the same time, the exact antisymmetry under exchanging any pair of fields, which follows from this term being topological, means there is no corresponding elastic channel $\chi_1\chi_1\to \text{SM}$, nullifying constraints from indirect and direct detection if the lighter of the two dark pions is taken to be the relic dark matter.
Rather, this mechanism predicts dark matter would be produced through novel channels {\em e.g.} $e^+e^-\to \gamma^\ast \to \pi_0 \chi_1 \chi_2$ (where the $\chi_2$ may be detector-stable or may decay at a displaced vertex) in colliders such as Belle II~\cite{Davighi:2024zip}, which will be further explored in the future.

Where do generalised symmetries come in?
The dark sector comes with a continuous $U(1)$-valued 1-form symmetry, thanks to there being a closed 2-form $\star j^{(2)}=\mathrm{Vol}_{S_2}$~\cite{Davighi:2024zjp}.
The topological interaction, for coefficient $n\in\Z$, has the effect of twisting the flavour symmetry of QCD with this 1-form symmetry into a 2-group generalised symmetry.
For a quick way to see this, consider turning on background gauge fields $A$ and $B$ for the 0-form symmetries and the 1-form symmetry respectively.
The minimal coupling of the 2-form background gauge field is 
\begin{equation}
    S\sim -\int_{\Sigma_4} B\wedge \mathrm{Vol}_{S^2}\, .
\end{equation}
The non-abelian version of \eqref{eq:2group} contains a particular 2-group transformation acting on the Goldstone and the 2-form field,
\begin{equation}\label{eq:Scoup}
    \pi_0\to\pi_0+f_\pi\alpha, \qquad B \to B + \alpha \frac{ne}{2\pi} F,  
\end{equation}
Precisely because the transformation of $B$ is twisted by the 0-form gauge parameter $\alpha$, the combination of \eqref{eq:topGauge} and \eqref{eq:Scoup} is invariant. In~\cite{Davighi:2024zjp} the full non-abelian 2-group structure is derived, including the modified Ward identities which encode the modified symmetry structure even with background gauge fields switched off.

The observation that this effective interaction encodes a 2-group structure constrains the possible RG flows that could end up on this EFT.
This `symmetry matching', generalising the more familiar tool of anomaly matching, was used in~\cite{Davighi:2024zjp} to identify a candidate completion of this EFT. A heavy abelian gauge field (whose field strength defines a 1-form symmetry current) couples to baryon number on the QCD side, which becomes identified with the topologically-conserved $\star j_B=\frac{1}{24\pi^2}\mathrm{Tr}(g^{-1}dg)^3$ in the chiral Lagrangian~\cite{Balachandran:1982dw,Witten:1983tx}, while the dark sector can be completed via a gauged linear sigma model. In the UV phase, the 2-group structure appears as an operator-valued mixed anomaly between the QCD chiral flavour symmetries and the heavy abelian gauge field.

\section{Non-invertible symmetries} \label{sec:NIS}

Finally, we conclude with a short introduction to a different kind of generalisation of symmetry, in which the group property satisfied by the algebra of topological operators is relaxed. 
This idea of {\em non-invertible symmetry} (NIS) has its origins in many examples from condensed matter and conformal field theory, the most well-known being Kramers--Wannier duality of the 1+1d critical Ising model (see {\em e.g.}~\cite{Shao:2023gho}).
NIS entails generalisation of the group multiplication law, $U_g(M_{d-1}) U_{g^\prime}(M_{d-1})=U_{gg^\prime}(M_{d-1})$ where $g,g^\prime\in G$, to a `fusion algebra':
\begin{equation}
    U_a (M_{d-1}) U_b (M_{d-1})=\sum_c N^c_{ab} U_c(M_{d-1})\, .
\end{equation}
Each $U_a$ need not have an inverse under the multiplication operation.
For example, the critical Ising model in 1+1d has symmetry operators $I$, $\eta$, and $\mathcal{D}$ that satisfy the algebra $\eta^2=I$, $\eta\cdot\mathcal{D}=\mathcal{D}\cdot\eta=\mathcal{D}$, and $\mathcal{D}\cdot \mathcal{D}=I+\eta$; the symmetry $\mathcal{D}$ has no inverse.
There are many such rich examples of NIS in lower-dimensional QFTs, but given our aim is to elucidate applications to particle physics we are interested primarily in 4d. 

\paragraph{Example: 4d theory with abelian ABJ anomaly.}

A class of examples of NIS in 4d is provided by chiral $U(1)$ gauge theories with a classical global $U(1)$ 0-form symmetry that suffers from an ABJ anomaly~\cite{Adler:1969gk,Bell:1969ts}. Let the dynamical field strength be $f=da$. 
The ABJ anomaly modifies the Ward identity for the global $U(1)$, from $d\star j=0$ to $d\star j=\frac{1}{16\pi^2} \mathcal{A} f\wedge f$ where $\mathcal{A}$ is the anomaly coefficient. Since $\star j$ is no longer closed, but rather its exterior derivative is equated to a non-trivial operator in the theory, na\"ively we have lost the topological property of our would-be defect operators $U_\alpha (M_{3})=\exp i\alpha \int_{M_{3}}\star j$. 
We might try to fix this up by shifting the definition of $\star j$ by an object whose derivative is $\propto f\wedge f$, which is precisely a Chern--Simons term. The would-be defect operator is then modified to $U_\alpha (M_{3})=\exp i\alpha \int_{M_{3}}\left(\star j- \frac{\mathcal{A}}{16\pi^2}  a\wedge f\right)$. But for a general angle $\alpha\in U(1)$, and allowing arbitrary topology for our (sub)-manifolds and bundles, the Chern--Simons contribution is not gauge-invariant and so this is not a well-defined fix.
Nonetheless, for rational angles $\alpha=p/q\in \mathbb{Q}$ ($p,q\in\Z$), one {\em can} give a proper gauge-invariant formulation of the Chern--Simons term by introducing an auxiliary gauge field on the defect; such {\em fractional Chern--Simons terms}, whose partition function we denote $Z_{\mathrm{CS}}^{p,q}$, appear in the effective action describing the fractional quantum Hall effect.
The upshot is that there do exist topological gauge invariant operators ({\em i.e.} symmetries) for each rational angle $\alpha=p/q$, but the composition of $U_{\alpha}$ and $U_{-\alpha}$ does not yield the identity but rather yields $Z_{\mathrm{CS}}^{p,q}(M_3)$. Hence the symmetry is non-invertible.

We can think of this family of non-invertible symmetry operators as being remnant symmetries that remain after the ABJ anomaly breaks the invertible global $U(1)$ symmetry of the classical theory. These symmetries remain unbroken even when the abelian gauge theory is put on spacetimes with non-trivial topology.

\paragraph{Example: pion decay.}
Of course, the chiral flavour symmetry of QCD with massless quarks, and with gauged QED, has ABJ anomalies. In the chiral Lagrangian, the WZW term mentioned above matches this anomaly. In this recasting via generalised symmetries, we learn that the coupling $S \propto n_c \int_{\Sigma_4} \frac{1}{8\pi^2}\pi^0 F \wedge F$ can also be understood to match the non-invertible symmetries that remain unbroken in the IR~\cite{Choi:2022jqy} -- similar to how the mixed WZW-like coupling \eqref{eq:topGauge} was seen to match 2-group generalised symmetry in an extension of QCD by dark pions on $S^2$.

\medskip

NIS provides a new class of possible unbroken symmetries with which we can understand selection rules in effective field theories. 
As for an ordinary symmetry, if we understand the ways in which a symmetry can be broken, then we have control over the size of such symmetry-violating perturbations.

There is no analogous NIS remaining in the related case of a $U(1)$ global symmetry that has an ABJ anomaly with a {\em non-abelian} gauge field. From this, we learn a way in which the NIS can be emergent in the IR, upon breaking a non-abelian gauge symmetry $G$ in the UV down to a gauged $U(1)\subset G$ in the IR which participates in an ABJ anomaly with a would-be global symmetry.

\paragraph{Example: gauge lepton-flavoured symmetries for tiny neutrino masses.}
This idea was put to work in~\cite{Cordova:2022ieu,Cordova:2022fhg} to explain how tiny neutrino masses might emerge from exponentially-suppressed effects coming from UV instantons that break NIS. The model invokes three right-handed neutrinos and a non-abelian gauged lepton number flavour symmetry $SU(3)_H$ in the UV, which has instantons, but is broken down down a gauged $U(1)_{L_\mu-L_\tau}$ (say) in the IR, for which the NIS structure is emergent. Yukawa couplings for Dirac neutrinos then follow the pattern $y_\nu = y_\tau \exp \left(-8\pi^2/g_H^2\right) \ll y_\tau$.

\paragraph{Example: Discrete NIS for flavour hierarchies.}
For our final example, we consider a theory in which a NIS arises {\em not} from an ABJ anomaly, but still has selection rules that have been shown to be of use in model-building. The setting is a compactification of Type IIB string theory on a 6-torus, which it is convenient to view as a product of three 2-tori, $X=T^2 \times T^2 \times T^2$.
Putting magnetic flux through these tori has the effect of (i) breaking the $U(1)$ translation symmetries down to discrete $\Z_N$ subgroups in the quantum theory, and (ii) forcing there to be chiral fermion zero modes acted on by these $\Z_N$ symmetries. The final ingredient of gauging a $\Z_2$ reflection symmetry turns these discrete $\Z_N$ symmetries into non-invertible symmetries~\cite{Kobayashi:2024cvp,Kobayashi:2024yqq}. The selection rules implied by these NIS, for different choices of $N$, give different `nearest-neighbour interaction' textures for the SM Yukawa matrices, providing a new playground for explaining quark and lepton masses and mixing angles.

\section{Outlook}

In the past decade, generalised symmetries have already taught us a huge amount about quantum field theories and string theory. In this talk we introduced the ideas of higher-form symmetry, higher-group symmetry, and non-invertible symmetry.
We are beginning to find interesting examples of all these generalised symmetries in particle physics, and this talk highlighted a small selection.
So far, many of the particle physics applications are in a sense
reframings of previously known phenomena. 
For instance, the consequences of ABJ anomalies have been understood more precisely via non-invertible symmetry, while axion shift symmetries have now been understood using 1-form global symmetry in 5d. 
There are, however, already examples of new phenomena and model-building ideas where generalised symmetries play a central role.
Moving to the future, it is reasonable to hope that generalised symmetries will find many applications in particle physics that tell us something completely new.

\paragraph{Acknowledgment.}

I am very grateful to the organisers of DISCRETE 2024, Ljubljana, for inviting me to give this plenary talk and for organising such an enjoyable and stimulating conference.

\bibliographystyle{JHEP}
\bibliography{refs}


\end{document}